\documentclass[conference]{IEEEtran}
\IEEEoverridecommandlockouts
\usepackage{cite}
\usepackage{amsmath,amssymb,amsfonts}
\usepackage{algorithmic}
\usepackage{graphicx}
\usepackage{textcomp}
\usepackage{xcolor}
\usepackage{booktabs}
\usepackage{tabularx}
\def\BibTeX{{\rm B\kern-.05em{\sc i\kern-.025em b}\kern-.08em
    T\kern-.1667em\lower.7ex\hbox{E}\kern-.125emX}}

\renewcommand{\thetable}{\arabic{table}}   
\renewcommand{\thefigure}{\arabic{figure}}

\makeatletter
\newcommand{\linebreakand}{%
  \end{@IEEEauthorhalign}
  \hfill\mbox{}\par
  \mbox{}\hfill\begin{@IEEEauthorhalign}
}
\makeatother

\begin{document}

\title{BigSUMO: A Scalable Framework for Big Data Traffic Analytics and Parallel Simulation\\
\thanks{Rahul Sengupta can be contacted at rahulseng@ufl.edu. The authors would like to thank Florida Department of Transportation (FDOT) for their support. Copyright notice: 979-8-3315-7391-1/25/\$31.00 ©2025 IEEE }
}

% \author{\IEEEauthorblockN{\textsuperscript{1}Rahul Sengupta, \textsuperscript{1}Nooshin Yousefzadeh, \textsuperscript{3}Yashaswi Karnati, \textsuperscript{1}Manav Sanghvi, \textsuperscript{1}Yash Ranjan, \textsuperscript{2}Jeremy Dilmore, \\ \textsuperscript{2}Tushar Patel, \textsuperscript{4}Ryan Casburn, \textsuperscript{1}Anand Rangarajan, \textsuperscript{1}Sanjay Ranka}
% \textit{\textsuperscript{1}University of Florida}, \textit{\textsuperscript{2}Florida Department of Transportation}, \textit{\textsuperscript{3}NVIDIA}, \textit{\textsuperscript{4}Kittelson and Associates}\\
% %Florida, U.S.A
% }

\author{\IEEEauthorblockN{Rahul Sengupta}
\IEEEauthorblockA{\textit{University of Florida} \\
Gainesville, U.S.A. \\
}
\and
\IEEEauthorblockN{Nooshin Yousefzadeh}
\IEEEauthorblockA{\textit{University of Florida} \\
Gainesville, U.S.A. \\
}
\and
\IEEEauthorblockN{ Manav Sanghvi}
\IEEEauthorblockA{\textit{University of Florida} \\
Gainesville, U.S.A. \\
}
\linebreakand
\IEEEauthorblockN{Yash Ranjan}
\IEEEauthorblockA{\textit{University of Florida} \\
Gainesville, U.S.A. \\
}

\and
\IEEEauthorblockN{Anand Rangarajan}
\IEEEauthorblockA{\textit{University of Florida} \\
Gainesville, U.S.A. \\
}
\and
\IEEEauthorblockN{Sanjay Ranka}
\IEEEauthorblockA{\textit{University of Florida} \\
Gainesville, U.S.A. \\
}
\linebreakand
\IEEEauthorblockN{Yashaswi Karnati}
\IEEEauthorblockA{\textit{NVIDIA Corp.} \\
Santa Clara, U.S.A.\\
}
\and
\IEEEauthorblockN{Jeremy Dilmore}
\IEEEauthorblockA{\textit{FDOT} \\
Deland, U.S.A. \\
}
\and
\IEEEauthorblockN{Tushar Patel}
\IEEEauthorblockA{\textit{FDOT} \\
Deland, U.S.A. \\
}
\and

\IEEEauthorblockN{Ryan Casburn}
\IEEEauthorblockA{\textit{FDOT} \\
Deland, U.S.A. \\
}
}

\maketitle
\begin{abstract}
% With increasing urbanization around the world, efficient management of urban traffic infrastructure is of vital importance to transportation agencies and city planners. It is thus vital to have a tool to help traffic authorities analyze large amounts of stored traffic data, as well as make impactful interventions to improve operations. Towards this end, we present BigSUMO, an end-to-end, scalable, open-source framework for analytics, interruption detection, and parallel traffic simulation.
% In this paper, we describe a system that ingests stored ATSPM data (high-resolution loop detector data and signal state data), along with sparse probe trajectory data. First, the system performs descriptive analytics as well as detects potential interruptions. The system is scalable to dozens of intersections, and can be used for both historical analysis and in real-time. Next, the system uses SUMO microsimulation for prescriptive analytics to test 100s of `what-if' counter-factual scenarios to optimize traffic performance. 

% The system is modular and can be modified to use different algorithms for data processing and outlier detection. We use open-source software and libraries to build the software pipeline for easy and cost-effective deployment. BigSUMO’s ability to simulate rich, scalable environments makes it an ideal platform for research and deployment in next-generation smart city mobility solutions, with potential applications in adaptive signal control, congestion pricing, and vehicle-to-infrastructure coordination.

With growing urbanization worldwide, efficient management of traffic infrastructure is critical for transportation agencies and city planners. It is essential to have tools that help analyze large volumes of stored traffic data and make effective interventions. To address this need, we present ``BigSUMO", an end-to-end, scalable, open-source framework for analytics, interruption detection, and parallel traffic simulation. Our system ingests high-resolution loop detector and signal state data, along with sparse probe trajectory data. It first performs descriptive analytics and detects potential interruptions. It then uses the SUMO microsimulator for prescriptive analytics, testing hundreds of what-if scenarios to optimize traffic performance. The modular design allows integration of different algorithms for data processing and outlier detection. Built using open-source software and libraries, the pipeline is cost-effective, scalable, and easy to deploy. We hope BigSUMO will be a valuable aid in developing smart city mobility solutions.

\end{abstract}


\section{Introduction and Related Work}
With the proliferation of traffic data collection sensors, Intelligent Transportation Systems (ITS) is transforming the way we manage and understand urban traffic dynamics. ITS integrates advanced technologies like multi-modal sensors (such as loop detectors, GPS, Connected Vehicle systems, video, LiDAR, etc.) into transportation infrastructure.

When looking at published research, we find analytics frameworks that focus on specific problems using different data modalities. \cite{midas}, \cite{vid1}, \cite{vid3} and \cite{sengupta2023towards} discuss end-to-end software pipelines that process traffic videos at intersections and present analytics with a focus on safety. %\cite{visualan} presents a visual analytics system for traffic congestion exploration and forecasting based on vehicle detector data, incorporating a Long Short-Term Memory (LSTM) model for congestion forecasting. \cite{visan2} presents a visual analytics framework for exploration of multidimensional road traffic data with a focus on anomaly detection. \cite{trajgraph} introduces `TrajGraph', a visual analytics method that works by creating a special graph to catalogue taxi trajectory data, thereby allowing the use of graph analysis algorithms. 
\cite{yashid} presents a loop detector-based interruption detection system at intersections, which detects traffic interruptions based on historic loop detector volumes. %\cite{telco} presents a system for road traffic understanding, management, and analytics using cellular phone metadata from cell towers operated by telecommunication operators. %\cite{survey1} presents a survey on data analytics applications in road traffic safety.

However, we find that there is a lack of a scalable end-to-end system that can process ATSPM and trajectory data at a city-wide scale, and provide descriptive and prescriptive analytics, using free and open-source components. In this work, we present ``BigSUMO: A Scalable Framework for Big Data
Traffic Analytics and Parallel Simulation" to address this deficiency. Our contributions include:

\begin{enumerate}
    \item We present an end-to-end system for analytics and simulation using ATSPM and sparse probe trajectory data.
    \item Descriptive Traffic Analytics, in the form of tables and figures, are presented in interactive Jupyter notebooks, which can be extended/customized as desired.
    \item Prescriptive Traffic Analytics, based on the results of running 100s of calibrated SUMO simulations, that can help investigate counterfactual scenarios and optimize traffic performance.
    \item We use open-source software components that can be used on any cloud platform running Python. We also use open GIS data published by public transportation authorities, such as the Florida Department of Transportation (FDOT).
    \item The system is parallel and scalable. The system is designed with minimal hardware and software dependencies and can be easily deployed by traffic agencies using commodity cloud services.
\end{enumerate}
%\footnote{AWS (aws.amazon.com), Azure (azure.microsoft.com), Google Cloud Platform (cloud.google.com)}

While ATSPM data is generally abundant, we acknowledge the present limitation of low Connected Vehicle penetration rate (between 3\% to 7\% of total traffic) in the data. In the near future, we expect Connected Vehicle penetration rates to rise, given the increased deployment of V2V/V2X infrastructure.

\begin{figure}[!htbp]
\centering
\includegraphics[width=0.45\textwidth]{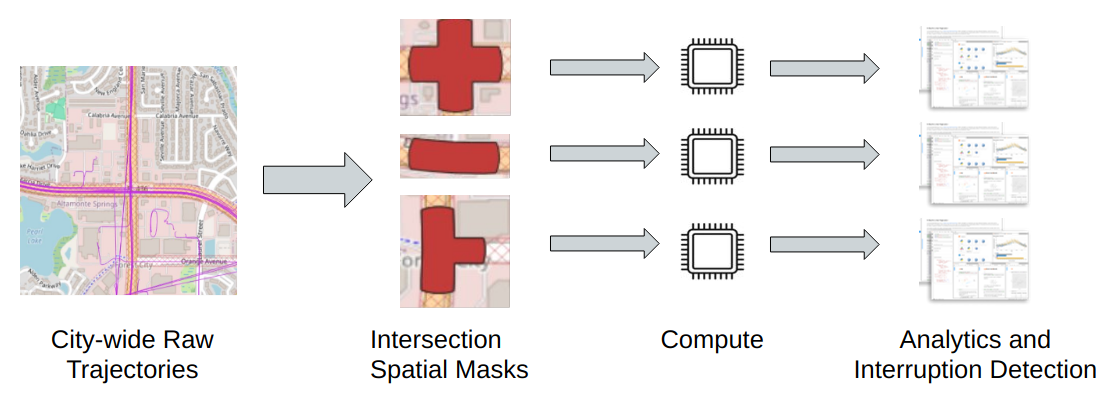}
\renewcommand{\figurename}{Fig.}
\caption{Use of intersection masks for Analytics and Interruption Detection}
\label{fig:masksoverview}
\end{figure}

% \bgroup
% \begin{figure}[!htbp]
% \centering \makeatletter\IfFileExists{pictures/ATSPMlog.pdf}{\includegraphics[width=0.35\textwidth]{pictures/ATSPMlog.pdf}}{}
% \makeatother 
% \renewcommand{\figurename}{Figure}
% \caption{A sample trace of high-resolution loop detector and signal state data. The data includes the ID of the signal, the timestamp, and codes describing the nature of the event. The data is stored in CSV format.}
% \label{ATSPMlog}
% \end{figure}
% \egroup

\section{Data Processing Pipeline}

ATSPM data obtained is already in a structured tabular format and requires minimal pre-processing. We filter the ATSPM data based on the intersection and time range required. %An example of ATSPM data format is show in Figure \ref{ATSPMlog}

On the other hand, the probe trajectory data needs to be processed. Our trajectory data processing pipeline directly ingests the raw data stored in Parquet or a similar cloud storage format. The dataset, obtained from a commercial vendor, contains between 17,000 to 22,000 unique journeys per day, spread over a large geographical region consisting of 3 counties in Florida, USA. Data is captured for two months in 2024. The data has a high temporal resolution of 3 seconds, with an approximate spatial accuracy of 1 meter. It is important to note that the penetration rate ranges between 3\% to 7\%. This reflects the proportion of all vehicles on the road at that time and location that were captured by the vendor's systems. The data collected is from regular passenger vehicles and not from a ride-share/taxi service. %Although the penetration rate is relatively sparse, the data collected still provides valuable insights. 
%\footnote{parquet.apache.org}

The trajectories themselves begin and end in locations such as parking lots, residential and commercial areas, etc. Driving behavior in these areas is very different from that observed on main roads. For the purposes of this work, we will only be concerned with roads that have been cataloged by the Florida Department of Transportation (FDOT) in the SunStore database. %The processes described below are applicable to datasets released by other state transportation authorities as well.
%\footnote{sunstore.cflsmartroads.com}

\subsection{Mask Generation Process}
In order to focus our data processing on relevant portions, we require roadway ``masks" in order to clip trajectories. Since it is tedious to draw masks by hand, we now describe a semi-automated process to generate masks. 

First, we download the “Basemap Route Roads” dataset from SunStore, which contains GIS (Geographic Information System) data for the roadways. It contains the geometric representation of the roadways in polyline-M format, for use with GIS software. We use QGIS to load and process this GIS dataset.
%\footnote{qgis.org}

Once loaded, we use the ``Buffer" function to create a 35-meter-wide buffer around the two lines that represent the 2 directions of the roadways. We merge their edges to form a contiguous, non-overlapping buffer that is 35 meters away from the two roadway lines.

Next, we use the “Intersections” dataset from SunStore to get the center locations of intersections. With this, we create circular buffers of 125 meters around each of the intersection center locations. We assume an intersection width of 50 meters, and an additional 100 meters on all sides to capture the region till the advanced loop detector (i.e. total of 125x2=250 meters diameter). We then use the ``Clip" function to clip the overlapping portions of the roadway buffers we generated earlier with the intersection buffers. %This gives us buffers that capture the various intersections as well as the road segments between them. The result of this can be seen in Figures \ref{fig:iscmask} and \ref{fig:segmask}.

% \begin{figure}[!htbp]
% \centering
% \includegraphics[width=0.45\textwidth]{pictures/iscmask.png}
% \caption{Examples of spatial masks for intersections.}
% \label{fig:iscmask}
% \end{figure}

% \begin{figure}[!htbp]
% \centering
% \includegraphics[width=0.45\textwidth]{pictures/segmask.png}
% \caption{Example of spatial masks for the road segment between intersections}
% \label{fig:segmask}
% \end{figure}

%With the masks obtained, it is possible to use them for trajectories queried at different time periods. The trajectories that are partially or wholly contained in the time period are filtered, and the masks are used to clip the trajectories spatially. We use GeoPandas \footnote{geopandas.org} and MovingPandas\footnote{movingpandas.org} libraries to perform this operation. 

\subsection{ Raw Trajectory Data Pre-processing}
The raw trajectory data is stored in multiple files across multiple directories, based on date and time. These files are combined into a single dataframe for further processing. The dataframe is then indexed and sorted by timestamp. 

We then filter out irrelevant and unreliable data points. We remove rows where the vehicle ignition status indicates the vehicle is off, as these points generally occur in parking lots, residential, and commercial complexes. They rarely occur on main roads. We remove journeys that do not meet a minimum duration threshold of 2 minutes.

%The cleaned data is converted into a geospatial format by associating latitude and longitude coordinates with each data point. We re-project the points from ``EPSG:4326" to ``EPSG:3857" in order to be able to work in meters.

The trajectories are then clipped using spatial masks (described earlier) to specific geographic areas of interest. This step focuses the analysis on relevant regions. Finally, we remove trajectories that are shorter than a specified minimum distance of 150 meters.

We use GeoPandas  and MovingPandas libraries to perform this operation. 

%\footnote{geopandas.org}
%\footnote{movingpandas.org}

%We use Pandas, GeoPandas, and MovingPandas libraries to perform various operations on the raw data. 

%In the next section, we discuss how we can generate useful traffic-related metrics to quantify and visualize traffic behaviors using the clipped trajectories. 

\section{City-scale Descriptive Analytics of Traffic States}

In this section, we describe various metrics we calculate based on the clipped trajectories. We then visualize the metrics using graph plots to gain better insights. We focus on intersection-level metrics, since intersections represent regions where various trajectories conflict, and are mediated by a traffic signal. Fig. \ref{fig:masksoverview} provides an overview.
%Traffic signals instruct vehicles to stop and proceed based on the traffic lights. Due to this, a wide variety of time-varying behaviors are seen.  Unlike segments that are sufficiently far away from the intersections where vehicles generally travel at the posted speed limit, regions near the intersection experience queuing. Because of this, vehicles must slow down and even stop, regardless of the speed limit. Braking behavior becomes important as well.

The analytics code is built using Pandas and presented in Jupyter notebooks. It can be used for any intersection of interest for the desired time period. We present an analysis for an hour of data collected at an intersection during weekday PM peak.

% \footnote{pandas.pydata.org}
% \footnote{jupyter.org}

% \subsection{Trajectory Length and Speeds}
% Speeds near intersections often vary greatly, on account of traffic signals and queueing. In Figure \ref{fig:lenspeeds}, we see histograms of trajectory lengths and the corresponding speed distribution seen during those trajectories.

% \begin{figure}[!htbp]
% \centering
% \includegraphics[width=0.45\textwidth]{pictures/lenspeeds.png}
% \caption{Histogram of trajectory lengths and average speeds (mph)}
% \label{fig:lenspeeds}
% \end{figure}

% Given that most vehicles go “through” (straight) the intersection, their trajectory lengths will be around 250 meters (i.e., twice 125 meters, which is the radius of the buffer). Other lengths include right-turns, which tend to be shorter, as well as lane changing behavior. We see that the speed distribution is reasonable, given that vehicles will often slow down and stop. The maximum speed seen is 50 mph, which is reasonable given the local speed limit.

\subsection{Wait Time and Locations}
Vehicles often have to wait at traffic intersections in response to a red light, as well as when approaching the end of a queue.

\begin{figure}[!htbp]
\centering
\includegraphics[width=0.35\textwidth]{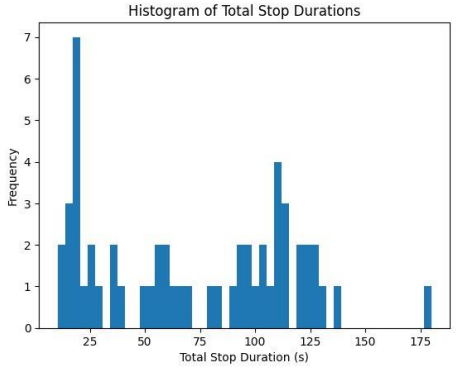}
\renewcommand{\figurename}{Fig.}
\caption{Histogram of vehicle stop times.}
\label{fig:stopdur}
\end{figure}

\begin{figure}[!htbp]
\centering
\includegraphics[width=0.35\textwidth]{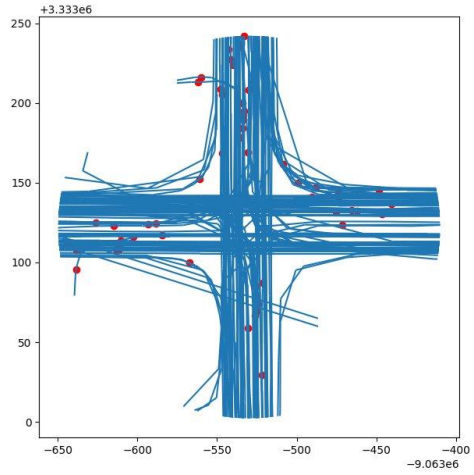}
\renewcommand{\figurename}{Fig.}
\caption{Plot of vehicle stop locations. Notice the two parallel trajectories moving from the left to the bottom-right. The vendor has hidden the intermediate trajectory points for privacy protection, but they are otherwise valid trajectories. We have included them in the plots in this analysis, but they can be removed if desired.}
\label{fig:stoploc}
\end{figure}

%Figure \ref{fig:stoploc} shows a plot of intersection trajectories with locations of stops. Note the two parallel trajectories moving from the left to bottom-right. The vendor has hidden the intermediate trajectory points for privacy protection, but they are otherwise valid trajectories. We have included them in the plots in this analysis, but they can be removed if desired.
Fig. \ref{fig:stopdur} shows a histogram of vehicle stop times.  This information is useful for signal timing optimization efforts, as lower wait times are more desirable. We see that the stop times are generally under 150 seconds, which is in the order of the cycle lengths seen at such intersections. 
In Fig. \ref{fig:stoploc}, we can see that the vehicles that were tracked often stopped well before the stop line. Given the sparseness of the data, it can be inferred that untracked vehicles (ahead of the ones that were tracked) had stopped in response to the traffic signal.

\subsection{Turning Movement Analysis}
An important benefit of having trajectory data is that the turning movement at the intersection is captured. This allows us to group trajectories based on which approach the vehicles came from and what turn they took at the intersection. Stop-bar loop detectors may not fully capture this information, especially when multiple turns are allowed for a particular lane (such as “through-right”). 

\begin{figure}[!htbp]
\centering
\includegraphics[width=0.35\textwidth]{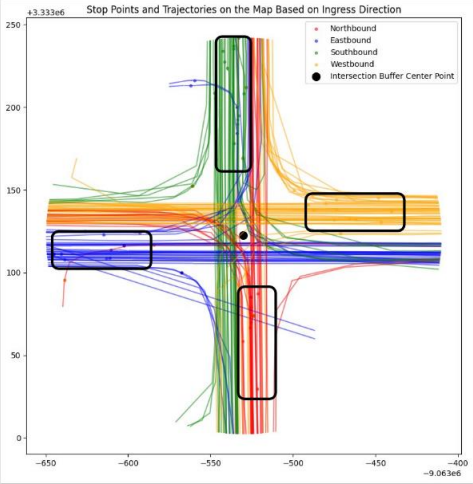}
\renewcommand{\figurename}{Fig.}
\caption{Approach-wise trajectories and stop locations.}
\label{fig:stopineg}

\end{figure}

\begin{figure}[!htbp]
\centering
\includegraphics[width=0.35\textwidth]{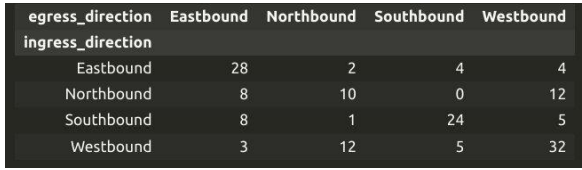}
\renewcommand{\figurename}{Fig.}
\caption{Approach-wise Origin-Destination matrix.}
\label{fig:odmatrix}
\end{figure}

\begin{figure}[!htbp]
\centering
\includegraphics[width=0.35\textwidth]{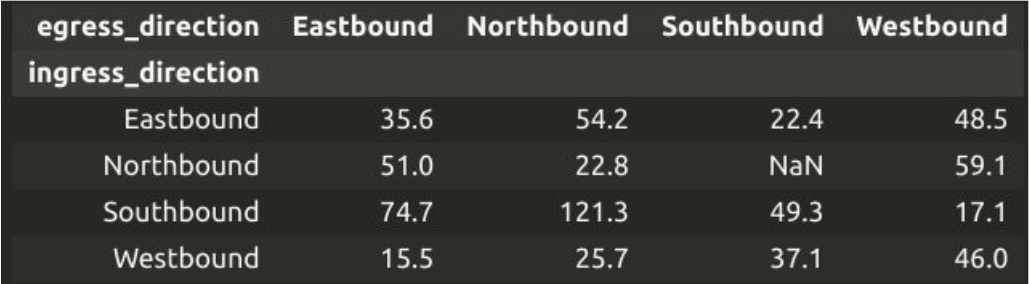}
\renewcommand{\figurename}{Fig.}
\caption{Table of approach-wise average travel times (s).}
\label{fig:avgtt}
\end{figure}

Fig. \ref{fig:stopineg} shows the plots of trajectories color-coded according to the approach they came from. The rectangular boxes show the locations of the incoming lane groups. Further, the data is tabulated in Fig. \ref{fig:odmatrix} in the form of an Origin-Destination matrix. This is a valuable input when running simulations of a single intersection\cite{karnati2024data}, as it allows the simulator to generate traffic for the intersection. Fig. \ref{fig:avgtt} shows the average travel time based on the origin and destination. This table provides valuable insight into which directions of travel need to be improved upon.

\begin{figure}[!htbp]
\centering
\includegraphics[width=0.35\textwidth]{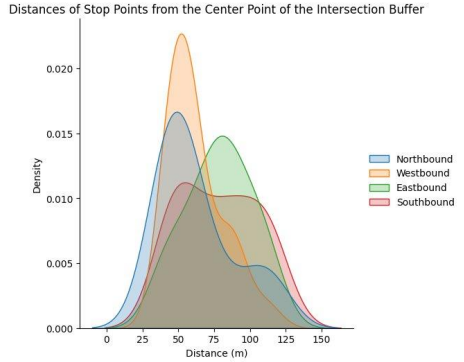}
\renewcommand{\figurename}{Fig.}
\caption{Approach-wise Queue length (m) probability distribution.}
\label{fig:qdistb}
\end{figure}

\begin{figure}[!htbp]
\centering
\includegraphics[width=0.35\textwidth]{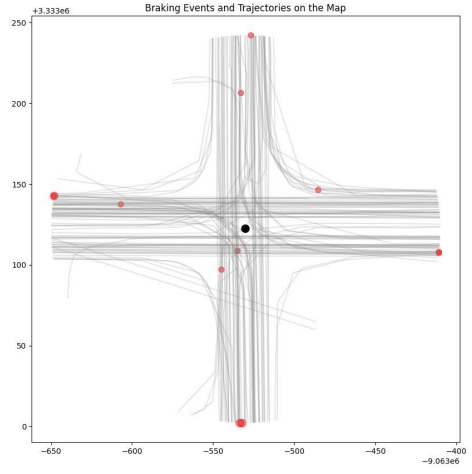}
\renewcommand{\figurename}{Fig.}
\caption{Locations where braking events of concern occurred.}
\label{fig:braking}
\end{figure}

\begin{figure}[!htbp]
\centering
\includegraphics[width=0.45\textwidth]{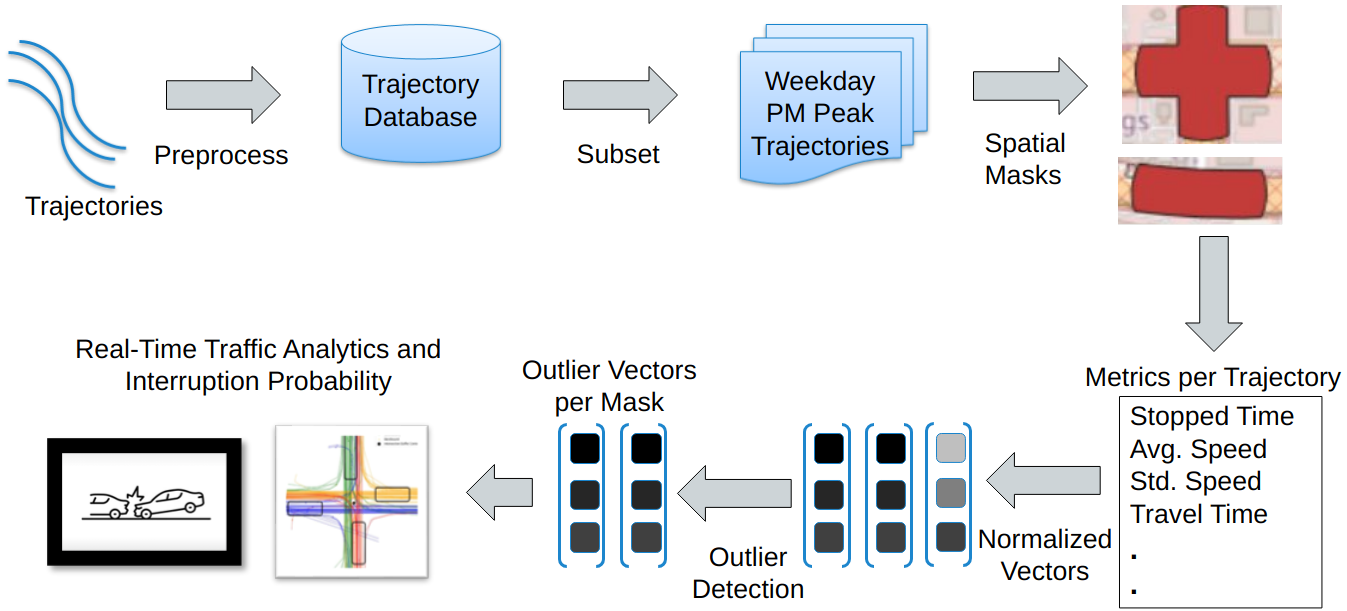}
\renewcommand{\figurename}{Fig.}
\caption{Overview of the Interruption Detection System.}
\label{fig:overview}
\end{figure}

\subsection{Queueing}
With the stopping information calculated from the trajectories, it is possible to infer the average queue length that a vehicle approaching the intersection from a particular direction may experience. We look for stops that are more than 10 seconds in length to determine where the vehicle entered the queue. %In Figure \ref{fig:qhist}, we can see a box-plot that shows the distance at which vehicles stopped from the center point of the intersection.

% \begin{figure}[!htbp]
% \centering
% \includegraphics[width=0.45\textwidth]{pictures/qhist.png}
% \caption{Approach-wise Stop location distances from the center of the intersection.}
% \label{fig:qhist}

% \end{figure}

 Using the data, we can fit a normal distribution to get a rough probability distribution of the queue length when a vehicle approaches the intersection along a particular direction. The distributions are plotted in Fig. \ref{fig:qdistb}.

It shows that Eastbound and Southbound directions have farther stop locations than Northbound and Westbound.

\subsection{Braking Behavior of Concern}
Since an intersection represents a zone of conflicting trajectories, it is possible that there may be dangerous situations such as near-misses. Unlike traffic collisions, these near-misses would generally not be captured in traffic records. However, we can use the trajectory data to look for braking behavior of concern. We use a braking threshold\cite{dsrcbraking} of –0.47g (where g is acceleration due to gravity) sustained for 2 seconds, to find locations where such braking occurs.

Fig. \ref{fig:braking} shows the locations of braking events of concern. Given the sparseness of the data, there are very few points. However, they still provide a general idea of where such braking is occurring.

\section{Interruption Detection Framework }
BigSUMO integrates two modules for detecting traffic interruptions, one using ATSPM data and another using trajectory analysis.  

The first interruption system, using ATSPM data, has been developed by the authors in a prior work\cite{yashid}, which relies only on loop detector data (from ATSPM). This loop detector-based system works by using granular phase-wise vehicle counts for a particular hour of the day, and compares them to the same hour the previous week and the week before that. Doing so allows the system to learn the general patterns that are expected at that hour of the day and the day of the week. 
%(Code available\footnote{github.com/yash-5595/interruptions-real-time-pipeline})

We now describe the second system, which uses trajectory data. As described in the previous sections, the pre-processed trajectories are stored in a database, and spatial masks are generated. A desired subset of the trajectory data is extracted from the database for outlier detection. We focus on trajectories recorded during weekday PM peak hours, which are critical for traffic analysis due to the high volume of vehicles. We apply spatial masks to the selected subset of trajectories, as described earlier. Various metrics are calculated for each trajectory within the spatial masks. Important metrics could include stopped time, average speed, speed variation (standard deviation), and travel time. These metrics provide quantitative measures to represent the behavior of trajectories. The calculated metrics are concatenated together and are normalized to create vectors. Each trajectory is thus represented by a vector. 
Additionally, we perform this process for the same spatial mask, at the same time of day and day of week, but for one and two weeks prior. This allows us to contextualize and infer what “normal behavior” of trajectories looks like. 

Outlier detection algorithms evaluate the normalized vectors to identify trajectories that deviate significantly from normal behavior. We use the Angle-Based Outlier Detection (ABOD) algorithm\cite{abod}. The algorithm works by looking at the variances of the angles between the difference vectors of data objects. The algorithm is fast and effective for high-dimensional data. We specify that the algorithm should consider 10 nearest neighbors for angle-based calculations for similarity.

We use PyOD library to perform the outlier detection. The code is flexible enough to use different features to generate vectors, as well as use different outlier detection algorithms. 
%\footnote{pyod.readthedocs.io/en/latest/}

The entire end-to-end process from trajectory clipping using spatial masks, pre-processing, generating analytics Jupyter notebooks, vectorizing, and performing outlier detection takes $\sim$2-3 minutes of wall-clock time per intersection. The process can be parallelized across intersections using libraries such as `multiprocessing'. This allows for the generation of analytics reports and interruption detection for multiple intersections using commodity CPUs with multiple threads.
%\footnote{docs.python.org/3/library/multiprocessing.html}

These outliers can indicate traffic interruptions or unusual events. The clustered outlier data can be used for downstream tasks such as generating a real-time traffic interruption probability heatmap. This heatmap visually represents the likelihood of interruptions across different areas and assists traffic authorities in detecting and responding promptly. Fig. \ref{fig:overview} shows an overview of the framework.

The system thus presented is modular and scalable. It can be used for multiple intersections in parallel and can use different outlier detection algorithms. At present, the ATSPM-based interruption detection system is much more reliable than the trajectory-based one. In the future, with the increasing penetration of Connected Vehicles, we may expect better performance with the trajectory-based system as well.

\section{Prescriptive Analytics using SUMO microsimulation}

In this section, we describe the use of the Prescriptive Analytics module that runs calibrated SUMO simulations in parallel, under a range of varying traffic parameters, and allows us to prescribe the best course of action to improve traffic performance. When deploying major traffic interventions (such as re-timing a corridor, modifying intersection geometry, etc.) in the real world, we would like to test out their potential impacts in simulation first. Ideally, we would want to test the robustness of the interventions under a wide range of likely parameters, such as traffic flow patterns, signal timing plans, etc. In order to facilitate this, we develop a parallel simulation model that uses multiple CPU cores to run the SUMO simulation with varying parameters. The system can easily run 100s of parallel simulations on commodity cloud infrastructure, without any license restrictions. We briefly describe the workflow here. For a more detailed treatment, please refer to \cite{karnati2024data}, Chapter 6.

\begin{figure}[!htbp]
\centering
\includegraphics[width=0.45\textwidth]{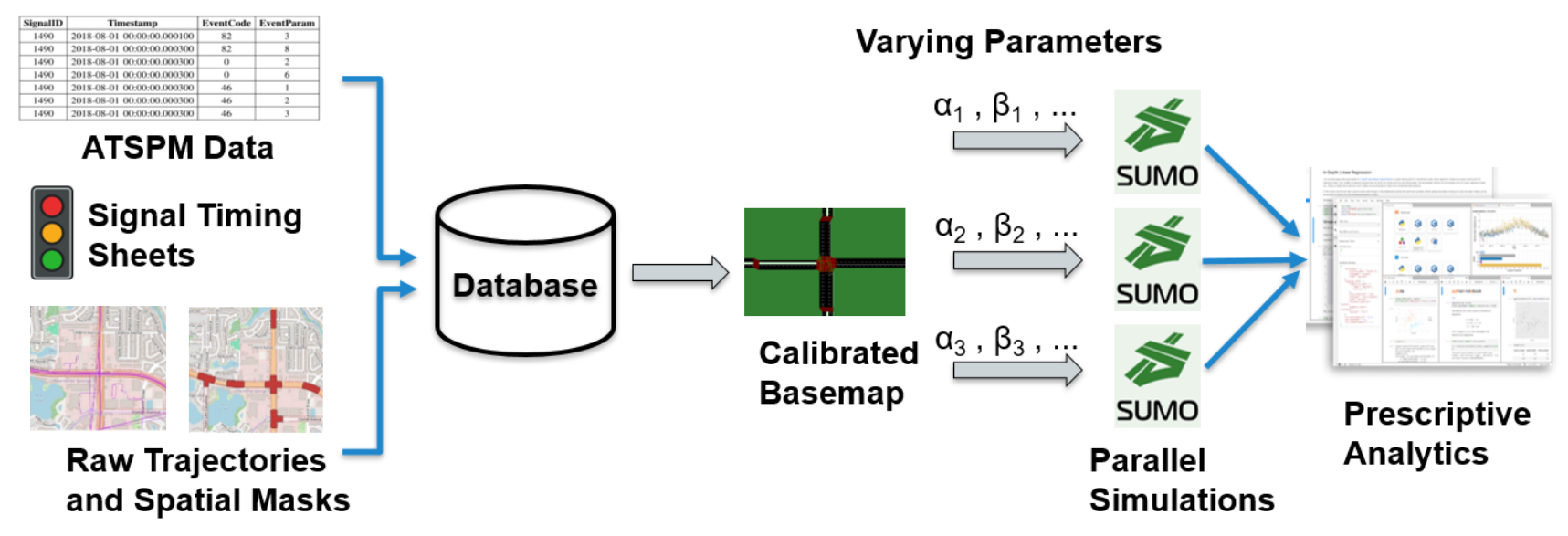}
\renewcommand{\figurename}{Fig.}
\caption{Overview of the SUMO-based Parallel Simulation Module.}
\label{fig:paralleloverview}
\end{figure}

We use Simulation of Urban Mobility (SUMO)~\cite{SUMO} microsimulation framework, since it is an open-source (GPL-licensed), highly portable, microscopic traffic simulator. It has been designed to handle large-scale road networks with heavy traffic, including vehicles and pedestrians. It has been widely used for research purposes. Fig. \ref{fig:paralleloverview} provides an overview of the system.
%\footnote{eclipse.dev/sumo/conference/}

\subsection{Basemap Calibration}
In order to simulate a portion of the road network, we need to build a basemap of the region of interest. We use the SUMO NETEDIT tool, a graphical network editor for manually editing road networks. Alternatively, the basemap can be imported from OpenStreetMap.
%\footnote{openstreetmap.org}

Once the basemap has been created, it is important to calibrate it to ensure it represents real-world conditions. If the field conditions are not appropriately replicated in simulation, any resultant metrics would be unusable. Various aspects of the simulation model need to be calibrated:

\subsubsection{Flow Calibration}
In order to calibrate simulated traffic flows, it’s necessary to ensure the correct number of vehicles begin and end at points that match actual observed movement. Loop detector data accurately counts vehicles at intersections, but it doesn’t reveal the details of their actual journey paths. By using the trajectory data, we can infer approximate origin-destination (O-D) matrices for various intersections, in the form of turning movement probabilities. We use the `routeSampler' tool to input vehicle counts (from ATSPM data) and O-D probability matrices. The output of this tool is a file that contains individual vehicles, with a predefined time when the vehicle enters the system. The route of the vehicle is pre-defined as well. This ensures the generated vehicles broadly follow the same trends in terms of vehicle counts and turning movement counts (TMC), as seen in the real world.
%\footnote{sumo.dlr.de/docs/Tools/Turns.html}

\subsubsection{Speeding Calibration}
Calibration must also take into account real-world speeding behavior, since SUMO’s default keeps vehicles at the speed limit. Trajectory data can reveal the distribution of maximum speeds observed at an intersection, especially after the signal turns green. This observed speed distribution is used to set SUMO’s `speedFactor', allowing vehicles to sample speeds according to a normal distribution with parameters derived from the observed distribution. This ensures simulated overspeeding matches observed behavior in the actual traffic data.

\subsubsection{Intersection Signal Configuration}
Our basemap consists of multiple intersections, each with four approaches, based
on standard NEMA (National Electrical Manufacturing Association) phasing. It consists of four through/right movements and four left-turn movements. At any time (generally), two of the eight non-conflicting movements have a green light, permitting safe traffic flow. The sequence of these green lights for phases is arranged in a sequence known as  `Ring-and-Barrier'. These values can be obtained from Signal Timing Sheets. If not, these values may be estimated by studying the ATSPM data to find maximum and minimum green times, and their orders. This information can be input to SUMO to program the traffic light behavior. 
Most intersections also have an exclusive left-turn buffer at each approach to accommodate left-turning traffic. This prevents the
left-turning traffic from blocking the through/right-turning traffic until the buffer is filled. The length of these buffers can be easily seen using an online mapping service such as Google Maps.

% \footnote{ops.fhwa.dot.gov/publications/fhwahop08024/chapter4.htm}
% \footnote{ops.fhwa.dot.gov/publications/fhwahop08024/chapter6.htm}
% \footnote{maps.google.com}

\subsection{Parallel Simulation}
Each simulation (simulating 1 hour) generally takes 3-6 minutes for a corridor with ~10 intersections, depending on the input traffic volume. If we wish to simulate a combination of varying parameters, such as flows, signal plans, etc., it could take several days to generate a large dataset. To speed up this process, we use Python's multiprocessing library to run dozens of simulations simultaneously, and store their results in XML format.

\subsection{Use Cases}
We describe use cases where BigSUMO was deployed for research. We intend to gradually clean, document, and release more of our source code (github.com/NSH2022/BigSUMO) in the future.

\subsubsection{Optimizing Signal Timing Plans}
An important application of having a parallel simulation framework is the ability to use grid search to find optimal signal timing plans for a corridor. For example, we can choose a set of viable parameters for various aspects of signal timing, such as common cycle length, minimum and maximum green times for the phases, etc. We can test them and find those parameters that yield an optimal result, often based on a performance metric like Corridor Travel Time. This aspect of BigSUMO can be compared to ReTime which uses VISSIM simulations in parallel. However, it is proprietary and requires licenses.
%\footnote{https://www.youtube.com/watch?v=SNrTXCOiWDQ}

\subsubsection{Data Generation for Deep Neural Models}
BigSUMO enables the large-scale data generation necessary for training and evaluating Deep Learning models. Especially, Graph Neural Networks (GNNs) have emerged as a powerful tool for learning spatial-temporal traffic dynamics using the inherently graph-structured nature of road networks with interdependent agents and properties. 

BigSUMO generates large-scale data, including granular lane-wise traffic data, traffic signal states, and vehicle behaviors under varying congestion scenarios. Moreover, BigSUMO’s ability to generate diverse and high-resolution sensor modalities allows researchers to construct multi-layered traffic graphs \cite{yousefzadeh2023comprehensive}. For example, \cite{10919173} uses Graph Attention Networks (GATs) to estimate lane-wise traffic waveforms time series at the location of exit and inflow loop detectors after being trained on graph data structured in a single or multiple layers. Also, ``Digital Twin" models emulating traffic behavior, such as MTDT \cite{10920162} and TGDT \cite{yousefzadeh2025tgdt} \cite{yousefzadeh2024dynamic}, provide comprehensive traffic performance evaluation and congestion prediction at urban traffic intersections and corridors. These works used data generated via BigSUMO for training and evaluation.

\section{Conclusion}
In this work, we developed an end-to-end system using ATSPM and sparse trajectory data for Descriptive and Prescriptive Analytics. We developed a data pipeline to ingest and process ATSPM and trajectory data. We then presented detailed analytics in the form of wait times and location maps, turn-movement ratios, braking severity maps, queuing distributions, etc. We also presented an interruption detection module based on the two data sources.
Next, we described how our framework allows us to investigate counterfactual scenarios under varying conditions, to study the potential impacts of traffic interventions. For this, we used a parallelized implementation of SUMO to simultaneously run multiple simulations. This system can also be used to generate data for training Deep Learning models. BigSUMO's potential applications include adaptive signal control, congestion pricing, and vehicle-to-infrastructure coordination.

\bibliographystyle{IEEEtran}
 \bibliography{bibliography}

\end{document}